\documentstyle[PASJadd]{PASJ95}

\newcommand{\vol}{50}
\newcommand{\no}{5}
\newcommand{\lett}{}
\newcommand{\spage}{???}
\newcommand{\rdate}{1998 January 14}
\newcommand{\adate}{1998 August 24}

\begin{document}
\setcounter{page}{1}

\title{ASCA Observations of the Seyfert 2 Galaxy NGC 7582: \\
       An Obscured and Scattered View of the Hidden Nucleus} 

\author{Sui-Jian {\sc Xue}$^{1,2}$, Chiko {\sc Otani}$^{1}$, 
Tatehiro {\sc Mihara}$^{1}$, Massimo {\sc Cappi}$^{1,3}$,
and Masaru {\sc Matsuoka}$^{1}$,
\\[12pt]
$^1$ {\it The Institute of Physical and Chemical Research (RIKEN), 
2-1 Hirosawa, Wako, Saitama 351-01, Japan}\\
{\it E-mail: xue@crab.riken.go.jp}\\
$^2$ {\it 
Beijing Astrophysics Center (BAC),
Beijing 100871, China}\\
$^3$ {\it Istituto per le Tecnologie e Studio Radiazioni Extraterrestri
(ITeSRE), CNR, iVia Gobetti 101, I-40129 Bologna, Italy}}

\abst{
We present the results of two ASCA observations of the Seyfert 2 galaxy
NGC 7582. The observation in 1996
revealed the variablity at $>$99\% confidence with
the shortest timescale of $\sim2\times10^4$ s. 
Variations were significant only in the hard X--ray band (2-10 keV) 
with the normalized variability amplitude $\sigma_{RMS}\approx 0.3$,
which is the same level as that of Seyfert 1 galaxies of the same luminosity.
In the soft X-ray band (0.5-2 keV) the flux stayed almost constant.
The source showed similar variability in another
shorter observation in 1994.

The overall broadband (0.5-10 keV) spectrum is complex. It shows
a heavily absorbed (N$_{\rm H}\sim1.0\times 10^{23}\rm\
cm^{-2}$) and flat ($\Gamma\sim 1.5$) power-law continuum plus
a slightly broad iron K$\alpha$ line at $\sim$ 6.4 keV with $EW\sim$ 170 eV.
Below 2 keV, a ``soft excess'' emission dominates the spectrum.

No significant variations in the average continuum and line fluxes  were
detected between the two observations. However, the inferred column density
increased by $\sim4\times10^{22}\ \rm cm^{-2}$ from 1994 to 1996.
This variation could be interpreted in terms of ``patchy torus'', namely 
suggesting that absorbing material on our line-of-sight comprises many
individual clouds. The observed iron line feature is also consistent with the 
picture of the transmission of nuclear X--ray continuum 
through the torus-like geometry. The nuclear X--ray luminosity
(2--10 keV) is $\sim 3 \times 10^{42}\ \rm ergs\ s^{-1}$,
similar to the typical values for Seyfert 1 galaxies.

The constant 0.5-2 keV soft X--ray spectrum,
though at least partially accounted for by some starburst contribution,
could be interpreted mostly as the scattered central continuum
from a spatially extended region. This is consistent with the lack of rapid
soft X--ray variability. Thus NGC 7582 is a typical example of a type 2 Seyfert
galaxy which exhibits both an obscured and scattered emission component
as expected from the AGN unification model.}

%
%
%

\kword{Galaxies: individual (NGC 7582) -- Galaxies: Seyfert 2 --
Galaxies: X--ray}

\maketitle
\thispagestyle{headings}

\section{Introduction}

NGC 7582 is a nearby ( $z=0.0053$ ) active galaxy in the southern sky, one of
the four spiral galaxies forming the Grus Quartet. Though it is categorized as
a Seyfert 2 galaxy according to its optical spectrum, 
it also joins a small class of so called ``narrow emission line galaxies'' (NELG)
which appear to have relatively bright and variable X-ray luminosity.
In a recent optical spectropolarimetry survey carried 
out by Heisler et al.\ (1997),
no scattered broad-emission lines were detected in NGC 7582. They interpreted 
this as seeing through an edge-on obscuring torus, providing that scattering 
particles polarizing broad-line photons lie very
close to or possibly within the plane of the torus.

In X--ray band, there are a number of studies on NGC 7582 based on previous 
space missions, e.g.\ HEAO-1/A2 (Mushotszky 1982), Einstein 
(Maccacaro \& Perola 1981; Reichert et al.\ 1985), 
EXOSAT (Turner \& Pounds, 1989), and more recent Ginga (Warwick et al.\ 1993).
These studies have suggested that NGC 7582 most probably contains
an obscured or highly absorbed (with column density $N_{\rm H} > 10^{23}\ 
\rm cm^{-2}$)
X-ray emission nucleus, like a typical type 2 Seyfert galaxy. However, 
some striking results were also revealed, especially that large factors of
variabilities in the absorption column density as well as in the continuum flux 
occurred during one or between some observations. These results could be 
the important key to investigate the nature of the absorbing matter along
the line-of-sight, as well as to test the AGN unification scheme in this 
special object. Unfortunately, these results are 
difficult to be interpreted, because several bright X--ray sources 
were detected by Einstein and ROSAT HRI in the immediate proximity of 
NGC 7582, including cluster S\'ersic 159-03 (Abell S1101),
active galaxies NGC 7590, NGC 7552
(Charles \& Philips, 1982), and a BL Lac object PKS 2316-423 
(Crawford \& Fabian 1994), which probably contaminated the former 
non-imaging X-ray observations. Therefore, to clarify the possible
variability properties of NGC 7582, multiple, spectral
imaging X-ray observations (with spatial resolution sufficient to separate out
the confusing sources in the field) are ideally required.

In this paper, we present the results of two ASCA observations of NGC 7582.
The imaging capability of ASCA clearly separated these 5 sources.
The broad energy band (0.5--10 keV) X--ray spectra obtained with ASCA 
provide a good opportunity to improve our understanding of 
this object significantly.  
We describe the observations and data reduction in section 2. The high 
quality data enables temporal and spectral analysis, 
which are presented in section 3 and section 4, respectively. The results are 
discussed in section 5. The conclusions of the paper are given in section 6.

\section{Observations and data reduction}

NGC 7582 was observed twice with the ASCA satellite. One observation  
was carried out on November 21-22, 1996 (AO4, hereafter), with Solid-state
Imaging Spectrometers, SIS-0 (S0) and SIS-1 (S1) operating in 1-CCD Faint
mode. The other one was made on November 14, 1994 (AO2 hereafter), 
with the SIS instruments 
operating in a mixture of Bright and Faint 2-CCD modes. For each  observation, 
Gas Imaging Spectrometers, GIS-2 (G2)
and GIS-3 (G3) were operated in normal PH mode. All data were selected 
from the intervals of high and medium telemetry rates. The SIS data were 
screened using the following criteria: a) The data was taken outside of
the South Atlantic Anomaly, b) the angle between the field of view
and the edge of the bright and dark earth exceeded 25$^{\circ}$ and 
5$^{\circ}$, respectively,
and c) the cutoff rigidity was greater than 4 GeV\ $c^{-1}$. After these
selections, we also deleted data if d) there were any spurious events
or the dark frame error was abnormal. For GIS data, no bright earth
angle and cutoff rigidity criteria were applied. More details concerning
the performance and instrumentation of ASCA  were reported in separate
papers (ASCA satellite: Tanaka et al.\ 1994; SIS: Burke et al.\ 1991;
GIS: Ohashi et al.\ 1996). 

The resulting effective exposures of
the observation in 1996 are $\sim 41.4$ks and $\sim 42.4$ks for the SIS and
GIS, respectively.  For the SIS data in mixture mode obtained in 1994,
the data was treated in BRIGHT (including converted FAINT data) mode, and
the resulting effective exposures of
the data are $\sim 22.5$ks for the SIS and $\sim 24.9$ks for the GIS. 

ASCA data of NGC 7582 did not suffer contaminations from any
confusing sources noted in section 1.
Source counts of both observations were extracted from circular regions 
of radius $\sim$ 3 arcmin and $\sim$ 5 arcmin for the SIS and GIS,
respectively. All the backgrounds were extracted from the source-free
regions in the same detectors. 

\section{Timing analysis}

Light curves of each observation were extracted
for all the four spectral instruments in XSELECT (V1.3),
with short (128 s) and long (5760 s, $\sim$1  satellite orbit)
binnings. Then the same
instrumental-type data, SIS0 and SIS1, GIS2 and GIS3 were combined
to improve the signal-to-noise ratio.
The resulting light curves show that the SIS and GIS data
are well consistent in each observation.

The variability properties were further investigated for
SIS light curves in the soft (0.5--2 keV) and hard (2--10 keV) energy bands
(figure 1). The result indicates that significant variation
($\sim$ a factor of 2) was only detected in
the hard X--ray band with typical timescale of
$\sim 2 \times 10^4$ s, which could be clearly seen on both 128 s and 5760 s
temporal analysis (figure 1). On the other hand, the soft X--ray
curve almost remained constant during each observation.
There was no sign of correlation between the hard and soft light curves.
According to the spectral study in the next section,
a break-point of the soft and hard components is $\sim 2$ keV.
Together with the tempotal analysis, we can conclude the existence of the
two independent spectral components in the spectrum of NGC 7582.

In order to check whether there was a possible correlated spectral
variability, such as a change of $N_{\rm H}$,
within the hard band during each observation, we have calculated the hardness
ratio HR (defined as the ratio of the count rate in 4--10 keV to the count rate
in 2--4 keV) for each bin of the light curve. The resulting HR is plotted
in figure 2 as a function of count rate for the SIS light curve of AO4. It
shows no significant spectral variability with the intensity.
The same result was obtained for AO2 data.

\begin{Fv}{1}
{5pc}
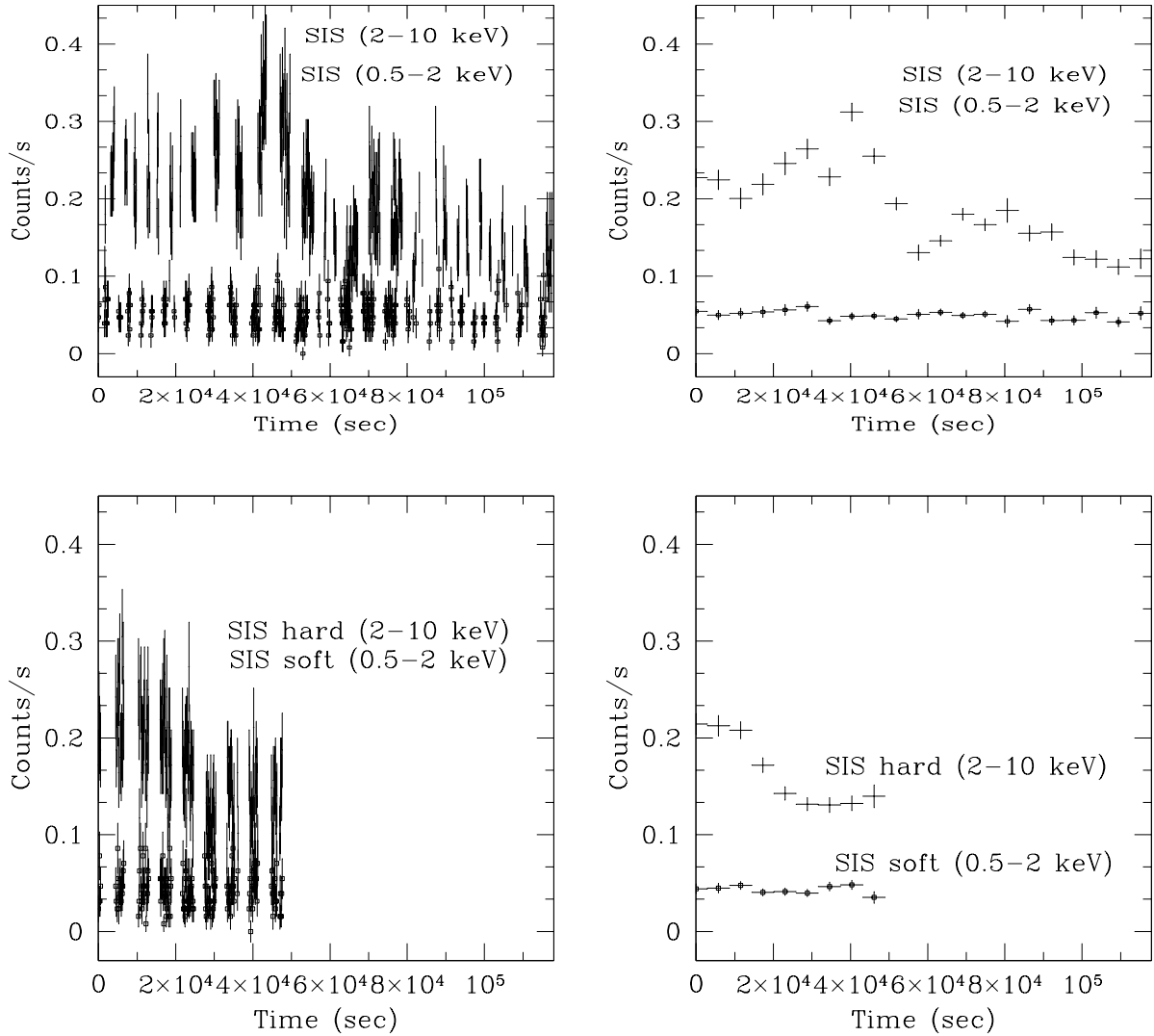
{The comparison of SIS hard and soft band light curves of NGC 7582.
Light curves in both short (128 s) and long (5760 s) bins are shown
for AO4 (upper panel) and AO2 (lower panel), respectively.
The time axis of AO4 starts from 18:24:13,
November 21 (UT), 1996; and AO2 starts from 05:47:32,
November 14 (UT), 1994.  For both observations,
the rapid variability is seen in the hard (2--10 keV) band, 
and no significant correlated variability
in the soft (0.5--2 keV) band. }
\end{Fv}

\begin{Fc}{2}
{5pc}
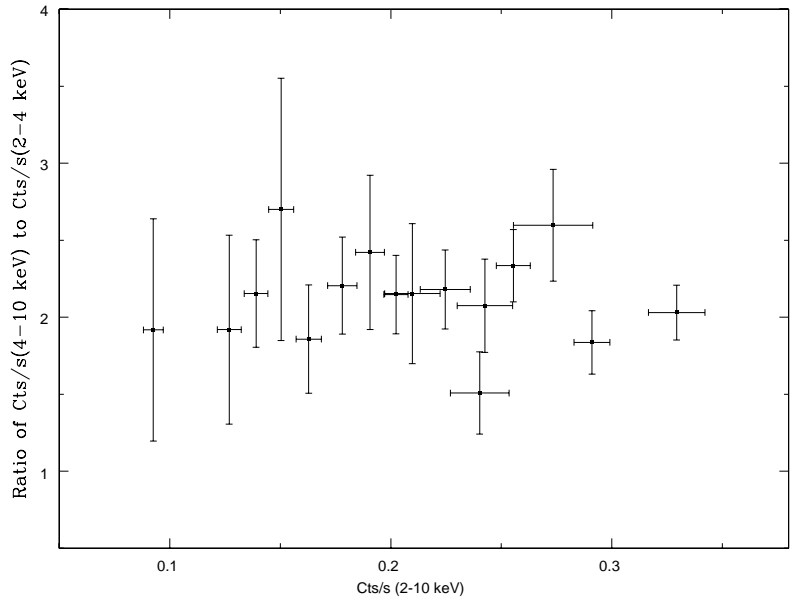
{The hardness ratio (4--10 keV)/(2--4 keV) as the function of SIS
count rate (2-10keV) in AO4. 
No spectral variability, such as a change of $N_{\rm H}$, was detected 
in the hard component.}
\end{Fc}

\section{Spectral analysis}

Since no significant spectral changes were detected in the individual soft
and hard spectral components in either observation,
we added together the data from each observation and
performed spectral fitting of the four time-averaged spectra simultaneously.
These spectra are: pairs of the SIS(0/1) in the energy of
0.5--10 keV, and the GIS(2/3) in 0.7--10 keV. 
The data was analyzed for each year; and 
independent normalizations were used for the SIS and GIS data,
because there are small uncertainties 
in the relative flux calibrations of different detectors.
The spectra were grouped so that each energy channel
contains at least 20 counts allowing minimization $\chi^{2}$
techniques. Spectral analysis have been performed
using XSPEC(V9.1) program (Arnaud et al.\ 1991).

Throughout this paper, the Galactic line-of-sight column density is fixed at
$N_{\rm Hgal}=1.47\times10^{20}\ \rm cm^{-2}$ (Stark et al.\ 1992). 
All errors reported below are quoted at the 90\% confidence level 
for one interesting parameter (i.e.\ $\Delta \chi^{2} = 2.7$). 

\subsection{Modeling of the soft excess component}

The broad-band spectra were initially modeled with a single absorbed 
power-law model.
Residuals of the fitting clearly indicate the presence of
significant ``excess'' emission that dominates the spectrum below $\sim$2 keV.
At higher energies, some small residuals around $\sim$6.4 keV
indicate the presence of an iron K$\alpha$ emission and/or an absorption
edge.


The poor description of the single power-law and the complexity of overall 
spectrum might be expected in the physical picture of unified AGN schemes. 
In the scheme, we might expect to see some fraction
of the primary continuum (power-law like) through the heavy obscuration of the
putative torus, with some components superimposed upon this, which could 
represent the indirect central continuum scattered
into our line-of-sight, and/or the leakage of the central continuum through 
a non-uniform (e.g.\ partial covering)
absorber. In addition, hot gas due to the starburst activity or in the host 
galaxy might also make an extra contribution to the soft X--ray emission. 
Thus, both the data and the possible physical picture lead us to explore some 
complex models for NGC 7582. 

We then attempt to fit the spectra by adding an extra spectral component 
to the single power-law to represent the upturn soft X--ray excess below 
$\sim$2 keV. The following continuum models were firstly investigated:


a) Scattering model. This model consists of the sum of two power-laws having the 
same photon index but different absorptions and normalizations. One power-law
is a direct component absorbed by a highly intrinsic column density,
and the other is a scattered component which is free from intrinsic 
absorption. The ``non-absorbed " scattered
continuum mimics the soft excess. This interpretation is 
supported by unified models of AGNs (Antonucci, 1993), which predicts that 
the direct continuum of Seyfert galaxies is strongly obscured by a thick torus
(Ghisellini, Haardt, \& Matt\ 1994), whereas a part of the nuclear emission is 
scattered by free electrons surrounding the broad-line region (e.g.\ Antonucci \&
Miller\ 1985; Awaki et al.\ 1991). The fraction of the scattered continuum
is evaluated by the ratio of the two power-law normalizations $A_s/A_p$.
The spectral fit of this model is shown in figure 3, and the best-fit 
parameters are listed in table 1.

It should be noted that the ``partial covering model'' is numerically 
equal to the scattering model. In this model,  
the ``non-absorbed'' power law component,
which might possibly explain the soft excess, 
physically represents a fraction ``leakage'' of the continuum from a partially 
obscured central X--ray source. 
In contrast to the scattered scenario, the soft component in the partial
covering model is expected to show the same 
rapid variations as the direct continuum.
From a correlation analysis of the soft (0.5--2 keV) with  hard (2--10 keV)
count-rate, the 1$\sigma$ upper limit of contribution from partial 
covering (i.e.\ the leakage of variable hard X--ray continuum) to the 
soft emission was not more than 10\%. 
Therefore, the partial covering model is immediately excluded by the data.
The constancy of the soft component 
is consistent with the scattering model because any rapid variations 
in the direct continuum would be smeared out in the emergent spectrum.

In the scattering model, the soft-emission is physically related to the 
hard-emission component; however there are also a few possibilities that
the soft emission is of thermal origin, and is completely independent on 
the hard-emission component. For example, b) Bremsstrahlung and 
c) Raymond-Smith model could account for the soft component.
In these cases, it is assumed the soft X--ray emission is produced 
by an optically thin  hot plasma associated with the host galaxy, 
e.g.\ starburst activities. The best-fit
of these two thermal models have minor difference, thus only the result 
of model c was quoted in table 1.

For each model we have also tested the significance for the presence of 
a narrow iron K$\alpha$ line, which was firstly fitted 
with free energy, and fixed width ($\sigma$). In all these 
models, thawing the line width could improve the line fitting at $>90$\%
confidence level ($\Delta \chi^{2} = 2.7$ for one interesting parameter) than 
that of narrow line fitting, we then let this parameter free.

It is clear from table 1 that both the scattering and thermal models give 
acceptable fits to 
the data with $\chi^{2}_{\nu} = 1.0 \sim 1.1$, 
and these models yield similar values of
column density, hard photon index and the similar iron K$\alpha$ feature 
in each observation dataset. Therefore, statistically these two type of 
models cannot be discriminated.
However, as we will discuss in section 5.3.1, the soft continuum in NGC 7582 
cannot be dominated by the thermal component. This suggests that the soft
continuum of the source is most likely dominated by 
the scattered component.  

\begin{Fc}{3}
{5pc}
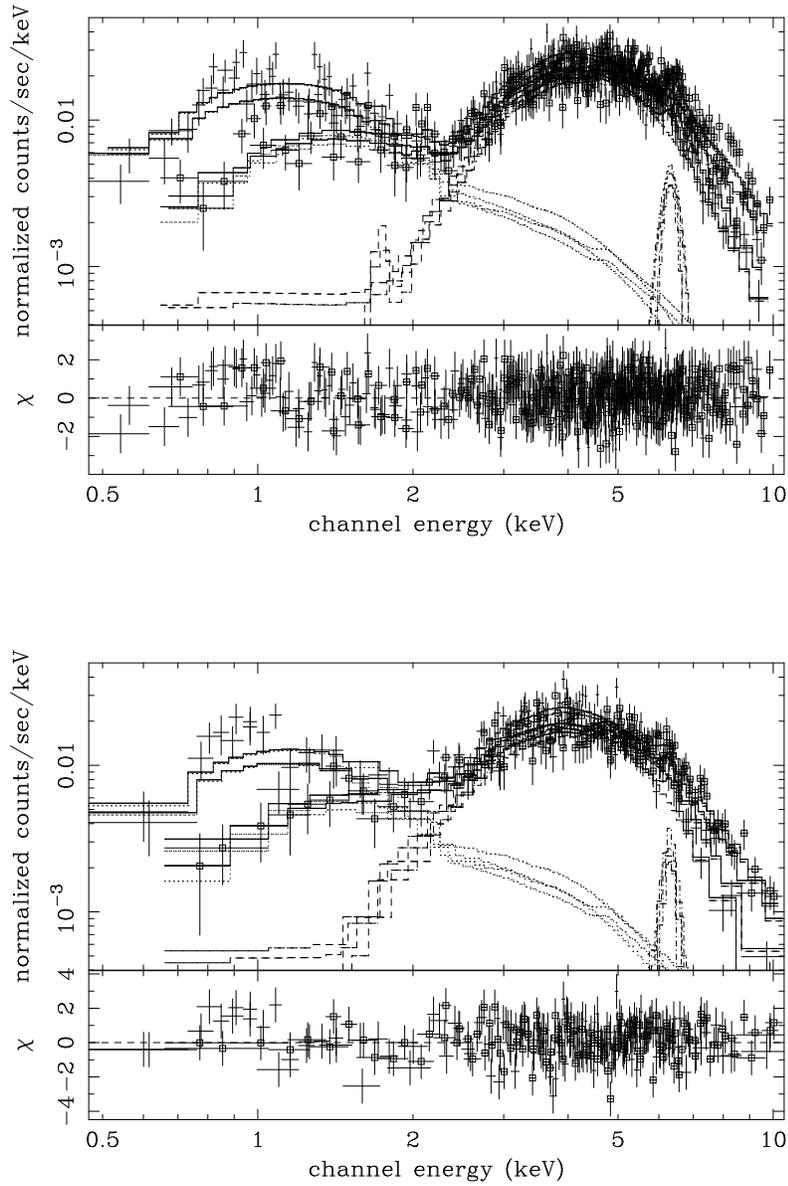
{The folded spectrum of NGC 7582, fitted simultaneously for
SIS(0/1) and GIS(2/3), with the scattering model.
Related residuals for AO4 (upper panel) and AO2
(lower panel) are also shown. The cross marker represents the SIS
data, and the void-square denotes the GIS data. 
The model gives acceptable description of the data
(see table 1), but still some residuals remain around $\sim$1 keV,
indicating that the soft spectrum is more complex than just scattered component.}
\end{Fc}

\subsection{The Best-fit Model}

The complexity of soft component is also suggested by the data.
Although the scattering model gives an acceptable fit of the data, it is 
not perfect. There are still some significant residuals 
below 1 keV (see figure 3), especially a small bump around 0.85 keV that might  
be of thermal origin, i.e.\ probably iron L-shell complex lines.
Therefore, we modeled this extra component using the
Raymond-Smith model, the best fit gives a the temperature of
$kT\simeq0.8$ keV, and abundance of $1.0-2.0$ solar (table 1). 
Although the abundance is poorly constrained
owing to the lack of unambiguous line features in soft X-ray spectrum,
the extra two free parameters improved the fit of the scattering model 
at $>99.9$ \% significance level with $\Delta \chi^{2} = 21$. 
The best-fit results are listed in table 1 and shown in figure 4.
In this complex model, the Raymond-Smith emission contributes at most $\sim 10-20$ \%
of the flux at 1 keV, which is consistent with the value 
expected from the FIR/X--ray luminosity correlation for normal and 
starburst galaxies (see section 5.3.1). 

\begin{table*}[t] 
\small
\begin{center}
Table~1.\hspace{4pt}{Spectral fit parameters}
\end{center}
\vspace{6pt}
\begin{tabular*}{\textwidth}{@{\hspace{\tabcolsep}
\extracolsep{\fill}}p{1pc}cccccccccc}
\hline\hline\\[-6pt]

{Data} &{$kT$} & {Abundance} & {$N_{\rm H}$} & {$\Gamma$}
& {$A_s/A_p^{a}$} & {$A_{RS}/A_s^{b}$}
& {$E_{K\alpha}$} & {$\sigma$} & {EW} & {$\chi^{2}_{\nu}$/d.o.f} \\
& [keV] & [solar] & [$10^{23}~{\rm cm^{-2}}$] & &\% &\% & [keV] & [keV] & [eV] & \\
[4pt]\hline\\[-6pt]

\\
\multicolumn{6}{l}{{\bf Scattering + Absorbed Power Law}~:}&\\ 
\\
1994 & - & - & $0.94^{+0.08}_{-0.08}$ & $1.49^{+0.14}_{-0.14}$ & $3.9^{+1.7}_{-1.2}$
& - & $6.31^{+0.09}_{-0.10}$ & $0.12^{+0.35}_{-0.07}$ &$170^{+80}_{-65}$ & 1.07/288 \\ 
\\
1996 & - & - & $1.32^{+0.06}_{-0.06}$  & $1.64^{+0.09}_{-0.09}$ & $3.2^{+0.9}_{-0.7}$
& - & $6.34^{+0.06}_{-0.06}$
& $0.15^{+0.11}_{-0.06}$ & $192^{+47}_{-46}$ & 1.05/578 \\
\\
\multicolumn{6}{l}{{\bf Thermal (RS) + Absorbed Power Law}~:}&\\ 
\\
1994 & $1.62^{+1.54}_{-0.51}$ & $<0.09$ & $0.75^{+0.05}_{-0.04}$
& $1.33^{+0.16}_{-0.16}$ & - & -& $6.30^{+0.09}_{-0.1}$ & $0.12^{+0.12}_{-0.08}$
& $170^{+70}_{-77}$ & 1.04/288 \\ 
\\
1996 & $2.48^{+1.40}_{-0.60}$ & $<0.08$ & $1.20^{+0.11}_{-0.08}$
& $1.52^{+0.13}_{-0.13}$ & - & - &
$6.33^{+0.07}_{-0.06}$ & $0.15^{+0.11}_{-0.06}$ & $180^{+53}_{-47}$ & 1.04/578 \\ 
\\
\multicolumn{6}{l}{{\bf Scattering + RS + Absorbed Power Law}~:}&\\ 
\\
1994 & $0.78^{+0.07}_{-0.12}$ & $2.6^{+1.6}_{-1.6}$ & $0.86^{+0.08}_{-0.08}$
& $1.38^{+0.12}_{-0.11}$ & $3.8^{+0.5}_{-1.5}$ & $ 16^{+7}_{-7}$
&$6.32^{+0.09}_{-0.10}$
& $0.12^{+0.16}_{-0.10}$ & $163^{+75}_{-66}$ & 0.99/286 \\
\\
1996 & $0.79^{+0.07}_{-0.09}$ & $1.4^{+0.5}_{-0.4}$ & $1.24^{+0.06}_{-0.08}$
& $1.52^{+0.09}_{-0.07}$ & $3.3^{+0.3}_{-0.4}$ &$14^{+5}_{-6}$ & 
$6.33^{+0.06}_{-0.0 5}$
& $0.15^{+0.10}_{-0.07}$ & $182^{+50}_{-40}$ & 1.02/574 \\ 
\\
[4pt]
\hline
\end{tabular*}
\vspace{6pt}
\noindent

$^{a}${$A_s/A_p=$ (normalization of the scattered component
at 1 keV)/(normalization of the primary continuum at 1 keV)}

$^{b}${$A_{RS}/A_s=$ (normalization of the RS component at 1 keV)/
(normalization of the scattered component at 1 keV)}

\end{table*}

\begin{table*}[t] 
\small
\begin{center}
Table~2.\hspace{4pt}{Flux and luminosity for the best-fit model}
\end{center}
\vspace{6pt}
\begin{tabular*}{\textwidth}{@{\hspace{\tabcolsep}
\extracolsep{\fill}}p{3pc}lccccc}
\hline\hline\\[-6pt]

{Dataset} & \multicolumn{3}{c}{Observed Flux [$10^{-13}\ \rm erg\ cm^{-2}s^{-1}$]}  
& \multicolumn{2}{c}{Intrinsic Luminosity$^{b}$ [$10^{40}~\rm ergs\ s^{-1}$]}  \\
[4pt]\cline{2-4} \cline{5-6}\\[-6pt]
&{0.5-2 keV} & {2-10 keV} & {2-10 keV$^{a}$} & {0.5-2 keV$^{c}$} & {2-10 keV$^{a}$} & \\
[4pt]\hline\\[-6pt]

1994 & $3.63^{+0.49}_{-0.67}$ & $151^{+36}_{-42}$ & $227^{+73}_{-54}$ &
$4.54^{+0.64}_{-0.71}$ & $273^{+89}_{-65}$ \\
\\
1996 & $4.18^{+0.34}_{-0.40}$ & $155^{+24}_{-22}$ & $272^{+46}_{-40}$ &
$5.06^{+0.66}_{-0.26}$ & $327^{+57}_{-46}$\\
\\
\hline
\end{tabular*}
\vspace{6pt}
\noindent

$^{a}${The absorption corrected.}
$^{b}${Assuming $H_0= 50~\rm km\ s^{-1}Mpc^{-1}$ and $q_0=0.5$. }
$^{c}${The Galactic absorption corrected. }
\end{table*}

\begin{Fc}{4}
{5pc}
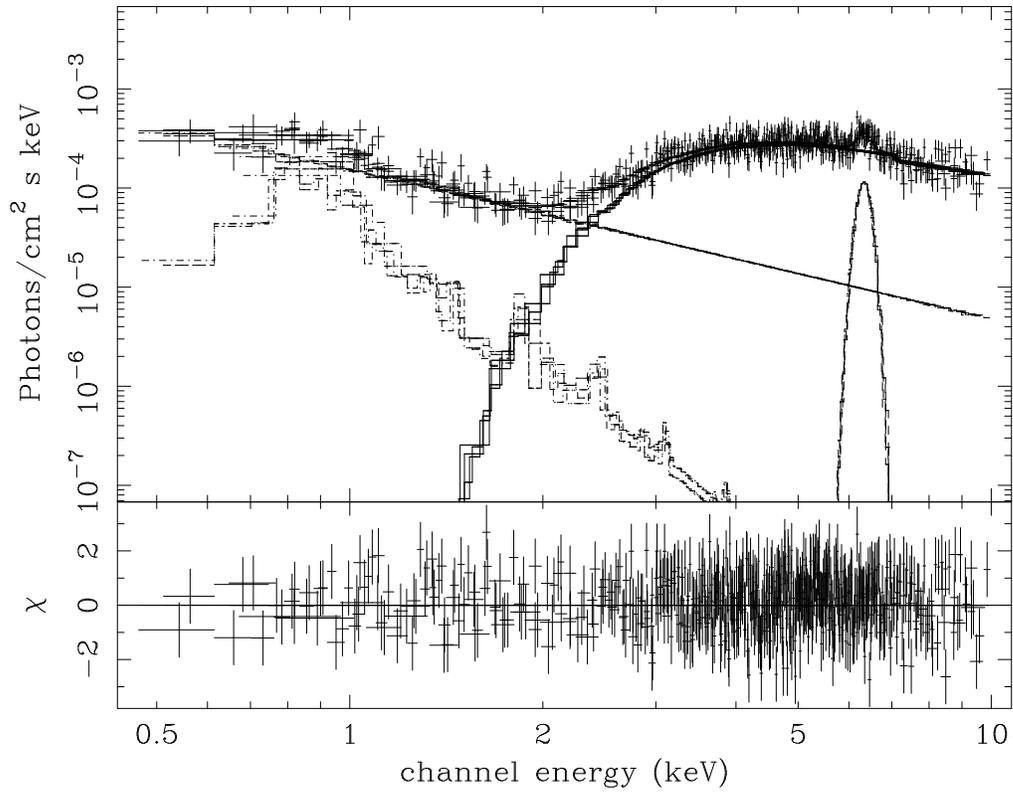
{The unfolded spectrum shown for AO4, fitted with the model of
``absorbed power-law + scattered power-law + Raymond-Smith +
Fe K line''. The best-fit parameters of this model and flux measurements
are listed in table 1 and table 2. }
\end{Fc}

Therefore, the plausible model which explains the data is that the soft 
emission consists of a mixture of the scattering component and the 
thermal component; the hard component is a heavily absorbed power-law 
continuum plus a slightly broad cold iron K$\alpha$ line. 
For this model, confidence contours of the photon index of the primary 
continuum vs.\ column density are
shown in figure 5. 

\begin{Fc}{5}
{5pc}
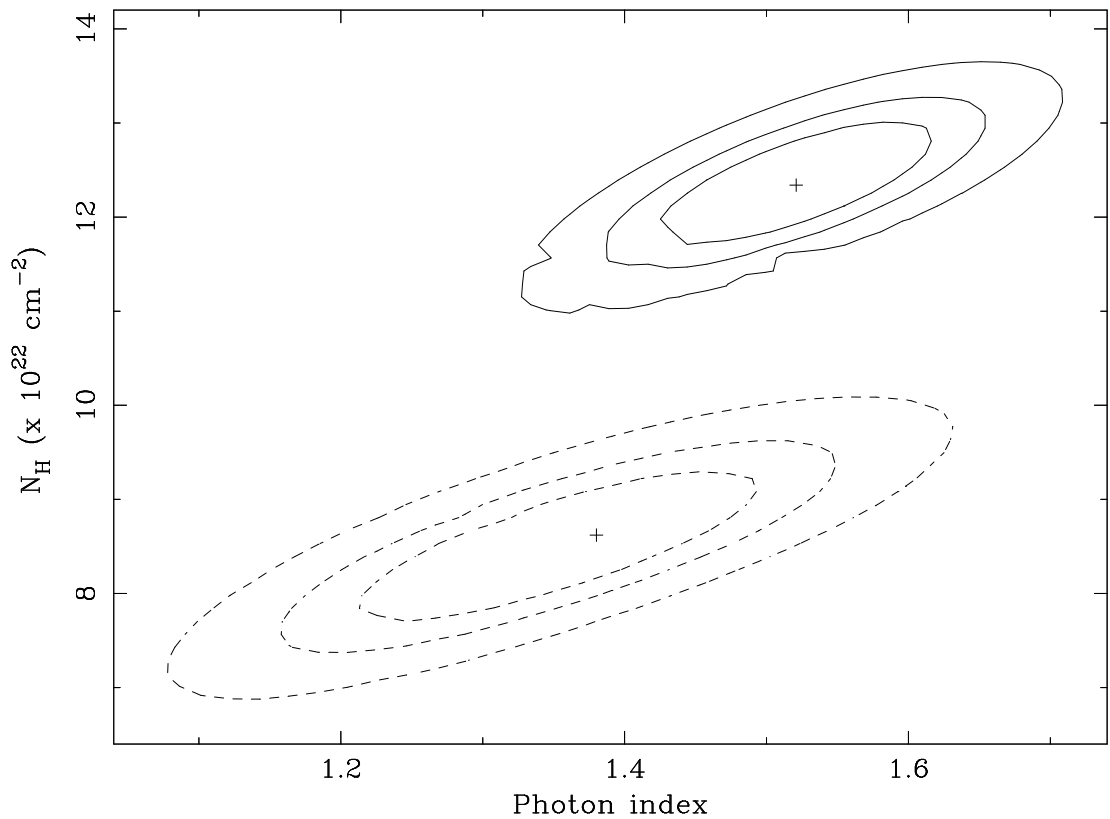
{Confidence contour levels (68\%, 90\% and 99\%) for the spectral slope
vs.\ the intrinsic absorption
column density. The solid and dashed lines are for AO4
and AO2 data, respectively. The photon-indices of the hard continuum are
in the range of 1.3-1.6 in the two
datasets. The intrinsic absorption changed evidently 44\% in the 2 year interval.}
\end{Fc}


Finally, we have searched for the presence of an iron edge in excess of that
modeled by the best-fit absorption in each dataset. Adding two extra
parameters (threshold energy and the depth of the edge) to the
best-fit model produced a slight improvement to the fit 
($\Delta \chi^{2} \simeq 4.6$) 
for AO4, and gave result of $\tau\simeq0.15^{+0.15}_{-0.10}$ at 
$\sim7.1 $keV. However, the AO2 data  
does not suggest any additional iron edge probably due to its lower
statistics. 
In this model, the excess absorption edge implies
either an iron overabundance of factor $\sim2$ or 
an additional X-ray absorption (partially covering) column in the line-of-sight
(see section 5.3.2).

\section{Discussions}
\subsection{Short-term variabilities during the observations}

Rapid X-ray variability is a very common property of Seyfert 1 galaxies, 
and the variability amplitude was found anti-correlated with the source 
luminosity (e.g.\ Nandra et al.\ 1997; Lawrence \& Papadarkis 1993). 
Therefore, to examine whether or not Seyfert 2 galaxies 
(for which we can detect their direct nuclear emission) also show the rapid 
variability, is a crucial test for the AGN unification scheme. 

Furthermore, it is widely examined that the complex X--ray spectra of type 2 
Seyferts in the ASCA energy band have different physical origins. To 
discriminate them, variability studies of various spectra components 
are important. 
NGC 7582 is a rare example of Seyfert 2 galaxies that clearly shows
rapid X--ray variability. Most compellingly, the hard and soft X-ray
emissions of this source have different
variability properties that indicate two distinct emission regions:
hidden nuclear hard X--ray emission and extended soft X--ray emission.

\begin{Fc}{6}
{5pc}
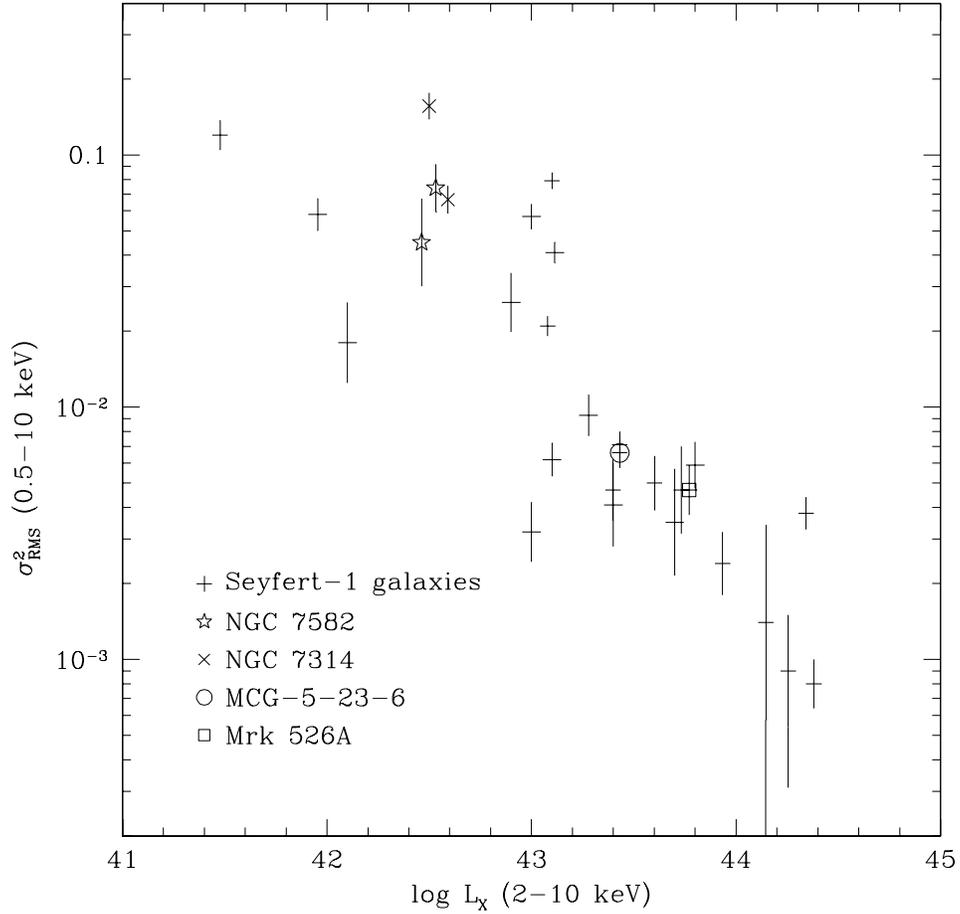
{Comparison of the rapid-variability characteristics of NGC 7582 seen in
2--10 keV with Seyfert 1 galaxies (Nandra et al.\ 1997), on the basis of 128 s
temporal analysis.  Also shown in the figure, three other type-2 objects:
NGC 526A, MCG-5-23-16 and NGC 7314 (Turner et al.\ 1997a).
For these Seyfert 2 galaxies, the rapid variability alone suggests
that the hidden Seyfert 1 nuclei are directly seen in the hard X--rays.  }
\end{Fc}

To test the idea that the observed highly variable continuum
comes from a hidden Seyfert 1 nucleus, we added our data points 
to the  $\sigma_{RMS}^2$ (0.5-10 keV) vs.\ $L_X$(intrinsic
luminosity in 2--10 keV) plot of Seyfert 1 samples  
(Nandra et al.\ 1997) (figure 6). The rapid-variability 
characteristics of the hard X--ray component in NGC 7582 at two different 
epochs are very similar to those of Seyfert 1 galaxies.
This suggests that the nuclear emission of NGC 7582 is similar to that of
Seyfert 1 galaxies. Turner et al.\ (1997a) also reported three other 
Type-2 objects, NGC 526A, MCG-5-23-16 and NGC 7314 that showed the same 
property as seen in figure 6. This result suggests that rapid 
X-ray variability is also common in Seyfert 2 galaxies. 

\subsection{Long-term variabilities between the observations}

Almost the same average 2--10 keV flux ($\sim1.5\times10^{-11}$ {$\rm erg\ cm^{-2}\ s^{-1}$}) 
was obtained in two ASCA observations. This flux is consistent with
the HEAO 1 result of $1.4\times10^{-11}$ {$\rm erg\ cm^{-2}\ s^{-1}$}
(Mushotzky, 1982), if assumed 
that the power-law component in their spectral model represents the source 
emission; and also consistent with the EXOSAT result of 
$1.7\times10^{-11}$ {$\rm erg\ cm^{-2}\ s^{-1}$}
(Turner \& Pounds 1989). Among the previous studies, 
Ginga observation reported the lowest
continuum level as $\sim0.6\times10^{-11}$ {$\rm erg\ cm^{-2}\ s^{-1}$}
(Warwick et al.\ 1993); and
that a substantial increase in N$_{\rm H}$ from $1.6\times10^{23}$ 
$\rm cm^{-2}$\ to 
$4.6 \times10^{23}$\ $\rm cm^{-2}$\ occurred in the $\sim$4 year interval between the 
EXOSAT and Ginga observations. All these results indicated that a complex long 
term spectral fluctuation might exist accompanying with a larger factor change 
in absorption column density.

The two ASCA observations separated 2 years provided a valuable chance to 
check for spectral variability over a longer time scales.
Comparing the contour plots of photon index vs.\ column density between the two
observations in figure 5, the spectral index does not show a significant difference.
However, the absorption column density evidently
increased by $\sim44$\% from AO2 to AO4,
confirming previous findings (Warwick et al.\ 1993). The present ASCA 
observations are amongst the best available in providing evidence for the 
variability of X--ray absorbing column within a Seyfert 2 nucleus, 
because most previous results (see Warwick et al.\ 1993) are based on a 
comparison of two measurements made by different instruments.

Variations in X--ray absorbing column are not common events in AGNs.
To our knowledge, so far it was only reported for a Seyfert 1.9 galaxy ESO103-G55
by Warwick et al.\ (1988) based on multi-observations made 
with EXOSAT. In that case, the significant variation in the column occurred over
a 90-day period, the author interpreted it in terms of an X--ray absorbing 
screen composed of broad-line clouds moving out of the line-of-sight.

However, the significant amount of X--ray absorbing column 
in type 2 AGN is generally associated with the putative obscuring torus. 
The partial obscuring of the nucleus of NGC 7582 would occur
in the outer region close to the edge of the torus (see section 5.4).  
If this region contains clumpy clouds 
(i.e.\ ``patchy torus''), its orbital
motion around the nucleus could cause the observed
variability of the line-of-sight absorption.

If this is the case, the transverse time scale, $D/v$, of a cloud with
the size, $D$, and the velocity, $v$, should be smaller than $\Delta t 
\sim 2$ years. $\Delta t$ should also be larger than 1 week, since
no variation of the column density was observed within one previous observation.
As the velocity is estimated by $(G M/R)^{1/2}$, where $R$ is the
distance from the nucleus and $G$ is the gravitational constant, this gives
\begin{equation}
\label{eq1}
D_{pc}\ \lower2truept\hbox{${< \atop\hbox{\raise4truept\hbox{$\sim$}}}$}\ 
\Delta t (GM/R)^{1/2} \sim  1.3 \times 10^{-3} M_8^{1/2} R_{pc}^{-1/2}, 
\end{equation}
\begin{equation}
\label{eq2}
D_{pc}\ \lower2truept\hbox{${> \atop\hbox{\raise4truept\hbox{$\sim$}}}$}\ 
1.3 \times 10^{-5} M_8^{1/2} R_{pc}^{-1/2}, 
\end{equation}
where
$M_8$ is the mass of the central blackhole in the unit of $10^8 M_\odot$.
$R_{pc}$ and $D_{pc}$ are in the unit of pc.

On the other hand, if the iron line comes from the same absorbing
matter, the ionization parameter 
$\xi \equiv L/nR^2$ of the matter should be $\xi<100\ \rm erg\ cm\ s^{-1}$,
since the energy of the
detected iron K line was 6.4 keV (see section 5.4) suggesting that the matter 
was cold or slightly ionized
(Kallman \& McCray 1982). Here $n$ is the density of the plasma.
Assuming the number of the clouds in the line-of-sight to be $N_c$,
the observed column density $N_{\rm H}$ was the sum the densities of 
the clouds, $ N_{\rm H} = N_c  n D$. 
Substituting  the source X-ray luminosity of $L = 3 \times 10^{43}\
\rm erg\ s^{-1}$ and $N_{\rm H} = 1 \times 10^{23} \rm cm^{-2}$,
\begin{equation}
\label{eq3}
D_{pc} < 1.0  R_{pc}^{2} N_c^{-1}.
\end{equation}
Eq.\ (\ref{eq1}) and (\ref{eq3}) give the upper limit of $D_{pc}$ as
\[
D_{pc} < 5 \times 10^{-3} M_8^{2/5}  N_c^{-1/5}.
\]
The maximum value is at $R_{pc} = 0.07 M_8^{1/5}  N_c^{2/5}$.\\
Eq.\ (\ref{eq2}) and (\ref{eq3}) give the lower limit of $R_{pc}$ as
\[
R_{pc} > 0.011 M_8^{1/5}  N_c^{2/5}.
\]
Since the dependencies of $M_8$ and $N_c$ are weak, 
$R_{pc}$ is consistent with the putative obscuring torus. 

Eq.\ (\ref{eq1}) and (\ref{eq3}) 
also give $D/R < 0.07 M_8^{1/5}  N_c^{-3/5}$, which  is consistent with the view that
there are many ($\sim 10$) small (compared with the size of the torus) clouds
near the edge of the torus.
The transverse motion of the clouds
could cause the observed change of the column density.
The number of the clouds in our
line-of-sight is large enough that NGC 7582 almost always shows the similar
absorption feature.

In the present case, the significant column variations
were observed on a timescale of years, therefore supporting the 
association of the X--ray absorbing matter with the putative torus. However,
more tight constraints on the physical conditions of the absorbing material 
can be expected from the future observations with time scales of 
week $\sim$ year.
 
\subsection{The spectral components}

\subsubsection{The origin of soft emission}

The lack of rapid variability in the soft X--ray band suggests that it has 
a different origin from the hard continuum. Some origins,
either electron scattering, thermal emission, or a mixture of the two, 
are expected to explain it. 

There are good evidence for starburst activities undergoing in the 
center region of NGC 7582. For examples, the optical spectrum of NGC 7582
shows high-order Balmer absorption features, which Ward et al.\ (1980) 
interpreted as evidence for a significant population of hot stars; also
its infrared spectrum is dominated by dust emission, and shows both dust
emission and silicate absorption features (Roche et al.\ 1984). However, we found 
that the thermal emission related to this starburst activities 
can not explain the total soft X--ray excess. According to the 
far-infrared(FIR)/X--ray luminosity correlation for normal and starburst 
galaxies (David, Jones \& Forman, 1992), we estimated, from the FIR 
luminosity of NGC 7582, the maximum expected starburst contribution
to the 0.5--4.5 keV luminosity is $\sim1.7\times10^{40}\rm\ erg\ s^{-1}$,
which is 3--5 factors below the derived luminosity of the soft X-ray
component in the thermal (e.g.\ Raymond-Smith) model (see section 4.1). 
Furthermore, the acceptable fits obtained with a RS thermal model need an 
abnormally low metal abundances, less than 0.08 solar values (see table 1),
this is unlikely to be real, because all known abundance gradients
in spiral galaxies suggest that the nuclei should be somehow enriched in
heavy elements with respect to solar (e.g.\ Vila-costas \& Edmunds, 1992).
There is one possibility to explain this low metal abundance,
i.e.\ the soft emission is a mixture of thermal component with
a featureless continuum, e.g.\ the scattered nuclear emission. However,
in this case, the thermal component with normal metal abundance
no more dominates the soft emission. All these results make the hypothesis of
a major thermal origin for the soft excess to be unlikely in NGC 7582.

As shown in section 4.2, the soft excess emission of NGC 7582 showed complex 
and composite spectrum, either the unique thermal or scattered origin can
be rejected by the data. We concluded that the soft spectrum is dominated by 
the scattered component with 10--20\% contribution from the starburst
emission. The ``scattering efficiency'', as defined by the normalization 
ratio of the scattered to the primary power-law continuum, was found to be of
3-4\%. Similar properties of the spectrum at low energies were also found
for many others Seyfert 2 galaxies (Ueno 1997; Turner et al.\ 1997a), with
the statistical distribution of scattering efficiencies in the range of 1--10\%.

The scattered view of the major soft X--ray emission of NGC 7582 was 
also expected by the presence of non-thermal UV emission (Kinney et al.\ 1991;
Mulchaey et al.\ 1992), which could not be seen through an absorption of 
$\sim 10^{23}$ $\rm cm^{-2}$. Thus it should be a scattered light of nuclear emission.  
One can expect that several soft emission lines will be produced in the warm 
gas which is responsible for the scattering, as seen in the ASCA spectra of
Mrk 3 (Iwasawa et al.\ 1994; Turner et al.\ 1997b).
It is, however, hard to discriminate such photoionized lines
in the present data, because of the poor statistics and contamination 
by the thermal starburst emission.

The present data showed evidently a scattered and obscured
view of a hidden Seyfert 1 nucleus. This naturally suggested that
a hidden broad line region (HBLR) should also be detected in the polarized 
spectrum in the context of the unified model for Seyfert galaxies.
However, Heisler et al.\ (1997) reported recently
a non-detection of polarized broad emission lines (PBLs) for
NGC 7582 in a well defined infrared-selected sample of Seyfert 2 galaxies.
On the one hand, there are difficulties in measuring weak 
PBLs, and we have little knowledge on how well  
the visibility of HBLR in Seyfert 2 galaxies is related to their
X--ray properties.

\subsubsection{The flat continuum}

The best-fit model for both observations require a strongly absorbed 
hard power-law continuum. Figure 5 shows that, within the 99\% error range, 
the two observations are consistent with a photon index of $\Gamma\sim1.3-1.6$.
This indicates that the primary X--ray spectrum of NGC 7582 is rather hard 
compared with the averaged slope ($\Gamma\approx 2.0$) found for most of 
other Seyfert 2s (Turner et al.\ 1997a). It is well established that presence
of reflecting matter in the nuclear region could strongly flatten the 
observed AGN continuum (Lightman \& White 1988; George \& Fabian 1991;
Matt Perola \& Piro, 1991), and Ginga observations of some Seyfert galaxies
supported this idea (Matsuoka et al.\ 1990; Pounds et al.\ 1990).

We have tried to add a reflection component (``plrefl'' model in XSPEC) 
in the best-fit model to test the possibility that the apparent flat 
spectrum could be due to an intrinsic steep continuum plus some cold 
reflection. The result (e.g.\ for AO4 data) showed that an intrinsic steep 
continuum with $\Gamma\approx1.85\pm0.28$ plus a certain reflection 
($R\approx0.88-4.22$) slightly improved the spectral fit at 
$\lower2truept\hbox{${> \atop\hbox{\raise4truept\hbox{$\sim$}}}$}$ 68\% 
confidence ($\Delta \chi^{2}=2$ with one additional free parameter).
Though the reflection ratio $R$, which describes the relative strength of
reflection component to the direct component, was not tightly constrained,
the present reflection model predicts an equivalent width of $250\pm150$ eV for
the iron line. This value is slightly larger than, but still consistent with 
the observed value of $EW\approx135 ^{+30}_{-90}$.

An alternative possibility to flatten a steep intrinsic continuum is through
the presence of an additional more strongly absorbed power-law. This scenario
usually be referred as ``Dual-Absorber'' model, which can be related to a 
physical scenario for type 1.5/2 AGNs in which the continuum source is 
completely covered by a torus of nonuniform density (see Weaver et al.\ 1994;
Hayashi et al.\ 1996). We found that a dual-absorber model
with an additional column of $N_{\rm H} \simeq 6.3^{+3.5}_{-2.8}
\times 10^{23}$ $\rm cm^{-2}$\ 
could explain the data with a steep power-law component similar to the case of
reflection model. The only constraint on such a complex model comes from
the iron K absorption edge at $\sim7.1$ keV expected by neutral absorption. The 
model predicts an absorption edge with optical depth $\tau\simeq0.16-0.33$ (90
\% confidence), which is still consistent with the measured value of 
$\tau=0.15^{+0.15}_{-0.10}$ (see section 4.2).   
Therefore, it seems that composite dual-absorber model is another plausible
explanation for the flat continuum of NGC 7582. The same model has also been
applied for several other sources observed with ASCA, like NGC 5252 
(Cappi et al.\ 1996), NGC 2110 (Hayashi et al.\ 1996) and IRAS 04575-7537 
(Vignali et al.\ 1998).

The remaining possible interpretation is that the spectrum
is intrinsically flat. This will pose a serious problem for theoretical models 
which are successful in the explanation of AGN's canonical X--ray slope at 
the value of $\Gamma \simeq 1.9-2.0$ (e.g.\ see Haardt \& Maraschi 1991; 1993).
There seems increasing number of Seyfert 2 galaxies with flat spectra revealed
by Ginga and ASCA, where Ginga data suggests that many of them can not be 
explained by the reflection model (Smith \& Done 1996); however, 
ASCA data suggests that many of them tend to be explained by the 
dual-absorber model. Actually, so far no any case could clearly claim 
to exclude both possibilities due to the limit of data quality. Therefore, 
whether or not there are Seyfert galaxies with intrinsic flat spectra is 
still an open question. 

\subsection{The iron K$\alpha$ line feature}

An iron K$\alpha$ line emission was significantly detected in both observations.
The best-fit line energy for both observations are consistent 
with the iron fluorescence line of 6.4 keV at rest from nearly-cold matter 
($<$ Fe XVI) (e.g.\ Makishima 1986). The observed equivalent width of 
$EW_{K\alpha}\simeq100-200$ eV is slightly larger than the prediction 
($\sim$70-90 eV) of the model where gas with 
inferred column density uniformly distributed around an isotropic X--ray source
(e.g.\ Inoue 1990). It, however, can be easily 
interpreted by the model that the surrounding matter is not uniformly
distributed, but in ``torus'' like geometry (Inoue et al.\ 1990; Awaki et al.\ 1991).
In such a case, since the observed absorption is Compton-thin,
and it is well believed that absorption tori in prototype Seyfert 2 galaxies 
are Compton-thick 
(i.e.\ $N_{\rm H} \lower2truept\hbox{${> \atop\hbox{\raise4truept\hbox{$\sim$}}}$} 10^{24} \rm cm^{-2}$),  
we probably observe the central source through the edge
of the torus. The observed complex configuration of X--ray absorbing
material along our line-of-sight, as suggested by its column variation 
as well as the sign of ``dual-absorber'', can be understood by such a scenario.

We note that observed equivalent width of iron K$\alpha$ is very similar to
the average result of $<$$EW_{K\alpha}$$>$ = $160\pm30$ eV found for Seyfert 1 galaxies
(Nandra \& Pounds 1994; Nandra et al.\ 1997). There are strong evidence that
iron K$\alpha$ lines in Seyfert 1 galaxies are originated from an accretion
disk near a supermassive black hole (Tanaka et al.\ 1995; Fabian et al.\ 1996).
For Type 2 AGNs, both the accretion disk in the inner region and the 
torus in the outer region are expected to be the source of the line emission. 
In fact, that line complex emitted from the two distinct X--ray reprocessors 
within a single type 2 AGN was observed by ASCA (Weaver et al.\ 1997). 

In the present case, the line width of $\sigma_{K\alpha}=0.15^{+0.10}_{-0.07}$
keV (FWHM$\sim9,000$--$28,000\rm\ km\ s^{-1}$) is significantly broader 
than the instrumental response ($\sigma\sim50$ eV
at 6 keV), but still much smaller than the mean width of 
$<$$\sigma_{K\alpha}$$>$ = 
$0.43\pm0.12$ keV found for the Seyfert 1 galaxies (Nandra
et al.\ 1997b). Therefore, it is unlikely that disk-line component is important
in NGC 7582. It could be the case that NGC 7582 has an edge-on  
accretion disk as generally expected for type 2 AGNs, and that the ultra-broad weak
line profile is smeared out with the continuum.

Line profile and variability studies provide determinative information on 
the origin of iron K$\alpha$ line. Unfortunately, such examinations for the
present datasets are inconclusive due to the limited energy resolution
of the line measurements. Much progress are expected to be made from the future
X--ray missions, such as AXAF, Astro-E and XMM.  

\section{CONCLUSION}

Two ASCA observations on the Seyfert 2 galaxy NGC 7582 in 1994 and 1996 yield
the following compelling results:

The hard X--ray (2--10 keV) flux was found highly variable on
time scales as short as 5.5 hours for both  
observations; meanwhile the soft band (0.5--2 keV) 
flux remained constant in both observations. The  
different variabilities in the soft and hard fluxes indicate
their different physical origins.
Specifically, the rapid variability amplitude of the hard X--ray flux
scaled by the source intrinsic luminosity are very similar to those of 
Seyfert 1 galaxies, implying it is the direct spectral component 
from the hidden Seyfert 1-like nucleus.

Time-averaged spectral studies revealed that the spectra, in the hard X--ray
(2--10 keV) band, are best described by a flat ($\Gamma\sim 1.3-1.6$) 
power-law continuum (i.e.\ the direct/primary continuum) absorbed 
by a column density of $0.8 \sim 1.3\times 10^{23}\rm\ cm^{-2}$. 
Though the slope is somewhat flatter than the average 
of other Seyfert 2 galaxies, the spectrum is consistent with a 
steep slope modified by a reflection or  complicated 
absorptions (e.g.\ ``dual-absorbers'') in the line-of-sight. 

A line feature was fitted with the center energy of $\sim6.4$ keV and 
the equivalent width of $\sim170$ eV, which corresponds
to the iron K$\alpha$ emission from nearly-cold matter.
This can be interpreted as the fluorescence emission 
from a non-uniformly distributed absorbing
material in a ``torus-like'' geometry. However, the slightly broad
line width (FWHM$\sim 15,000\rm\ km\ s^{-1}$) could also be a 
sign of unresolved multi-origins for the observed iron K line.
Future X--ray space missions with high energy resolution at this band
could hopefully determine the line emission regions.

The soft X--ray (0.5--2 keV) emission in excess of the absorbed direct
power-law continuum demonstrates a composite spectrum.
The dominant component is most likely due to scattered emission
from the nuclear continuum; the remaining contribution ($\sim20$\%) at
this band comes from a starburst component with temperature of 
$kT\approx0.8$ keV. The lack of any rapid variability in this band 
supports this interpretation. Such a scattered view of the soft
X--ray emission is also predicted for the Seyfert 2 galaxies
in the unified AGN model.   

A significant variation ($\sim 4\times 10^{22}\rm\ cm^{-2}$) 
in the absorption column 
was detected between the two observations in the 2-year interval. 
The variation timescale of $\sim$years might indicate that 
the X--ray absorbing material might be a ``patchy--torus'' composed of 
many individual clouds. 

The measured column density implies the ``torus'' is 
Compton-thin. This might indicate that either this object lies at an 
intermediate inclination angle between Seyfert 1's and a prototype 
Seyfert 2's, or the obscuring torus is not always Compton thick. 
In any case, NGC 7582 is one of the typical 
examples of the Seyfert 2 galaxies which exhibit both obscured 
and scattered emission components in X--rays as predicted by 
the unification scheme for Seyfert galaxies. In many details, the ASCA 
observations of this source provided very good tests for the unified models.
\par
\vspace{1pc} \par
We thank all the members of the ASCA team who operate the satellite and maintain
the software and database. 
This research has made use of the NASA/IPAC Extra-galactic Database (NED)   
which is operated by the Jet Propulsion Laboratory, California Institute   
of Technology, under contract with the National Aeronautics and Space      
Administration. S.J.X. acknowledges the financial support from an exchange
program between the Chinese Academy of Sciences (CAS) and RIKEN, and a partial
support from the President Foundation of CAS.
S.J.X. is currently belongs to BAC which is jointly sponsored by the 
Chinese Academy of Sciences and  Peking University.

\newpage
\noindent

\section*{References}
\small

\re Antonucci R.R.J.\ 1993, ARA\&A  31, 473
\re Antonucci R.R.J., Miller J.S.\ 1985, ApJ  297, 621
\re Arnaud K.A. et al.,\ 1991, XSPEC User's Guide(ESA TM-09)
\re Awaki H., Koyama K., Inoue H., Halpern J.P.\ 1991,
                 PASJ  43, 195
\re Burke B.E., Mountain R.W., Harrison D.C., Bautz M.W.,
                 Doty J.P., Ricker G.R., Daniels P.J. 1991, IEEE Trans. 
                 ED-38, 1069
\re Cappi M., Mihara M., Matsuoka M., Brinkmann W., Prieto M.A,,
                 Palumbo G.G.C.\ 1996, ApJ  456, 141
\re Charles P.A., Philips M.M.\ 1982, MNRAS  200, 263
\re Crawford C.S., Fabian A.C.\ 1994, MNRAS  266, 669
\re David L.P., Jones C., Forman W.\ 1992, ApJ  388, 82
\re Fabian A.C., Nandra K., Reynolds C.S., Brandt W.N.,
                Otani C., Tanaka Y., Inoue H., Iwasawa K.\ 1995, MNRAS  277, L11
\re George I.M., Fabian A.C.\ 1991, MNRAS  249, 352
\re Ghisellini G., Haardt F., Matt G. 1994, MNRAS  267, 743
\re Hayashi I., Koyama K., Awaki H., Ueno S., Yamauchi S.
                 1996, PASJ  48, 219
\re Heisler C.A., Lumsden S.L., Bailey J.A.\ 1997, Nature  385, 
                 700
\re Lawrence A., Papadakis I.E.\ 1993, ApJ  41, L93
\re Leighly K.M., O'Brien P.T.\ 1997, ApJ  481, L15
\re Lightman A.P., White, T.R.\ 1988, ApJ  335, 57
\re Kallman T.R., McCray R.A.\ 1982, ApJs  50, 263
\re Inoue H.\ 1990, in Proc. 23rd ESLAB Symp., ed. J. Hunt,
                 B. Battrick, Paris: Eur. Space Agency. p783 
\re Iwasawa K., Yaqoob T., Awaki H., Ogasaka Y.\ 1994, 
                 PASJ  46, L167 

\re Maccacaro T., Perola G.C.\ 1981, ApJ  246, L11
\re Makishima K. 1986, in The Physics of Accretion onto Compact
                 Objects, ed. K. O. Mason, M.G. Watson \& N.E. White
                 (Berlin: Springer), 249
\re Matsuoka M., Yamauchi M., Piro L., and Murakami T.\ 1990, 
                 ApJ  361, 440
\re Matt G., Perola G.C., Piro L.\ 1991, A\&A  247, 25
\re Mulchaey J.S., Mushotszky R.F., Weaver K.A.\ 1992, ApJ  390, L69
\re Mushotszky R.F.\ 1982, ApJ  256, 92
\re Nandra K., Pounds K.A.\ 1994, MNRAS  268, 405
\re Nandra K., George I.M., Mushotzky R.F., Turner T.J. \& 
                 Yaqoob T.\ 1997a, ApJ  476, 70
\re Nandra K., George I.M., Mushotzky R.F., Turner T.J. \& 
                 Yaqoob T.\ 1997b, ApJ  477, 602
\re Ohashi T., Ebisawa K., Fukazawa Y., Miyoshi K., Horii M,
                 Ikebe Y., Ikeda Y., Inoue H., et al. 1996, PASJ 48, 157
\re Osterbrock D.E.\ 1989, in {\it Astrophysics of Gaseous Nebulae
                 and Active Galactic Nuclei}, (San Francisco: Freeman)
\re Polletta M., Bassani L., Malaguti G., Palumbo G.G.C., 
                 Caroli E.\  1996, ApJS  106, 399
\re Pounds K.A., Nandra K., Stewart G.C., Geoge I.M.\ 1990, Nature  
                 344, 132 
\re Reichert G.A, Mushotszky R.F., Petre R., Holt S.S.\ 1985, ApJ 
                 296, 69
\re Roche P.F., Whitmore B., Aitken D.K., Phillips M.M.\ 1984, 
                 MNRAS  207, 35
\re Serlemitsos, P., Ptak, A., Yaqoob, T.\ 1997, in The Physics
                 of LINERs in View of Recent Observations, ed. M. Eracleous,
                 A. Koratkar, C. Leitherer, \& L. Ho, 70
\re Stark, A.A., Gammie, C.F., Wilson, R.W., Bally, J., Linke, R.A.,
                 Heiles, C., Hurwitz, M.\ 1992, ApJS, 79, 77
\re Storchi-Bergman T., Kinney A.L., Challis P.\ 1995, ApJs  
                 98, 103
\re Smith D.A., Done C.\ 1996, MNRAS  280, 355
\re Tanaka Y., Inoue H., Holt S.S. 1994, PASJ 46, L37
\re Tanaka Y., Nandra K., Fabian A.C., Inoue H., Otani C.,
                 Dotani T., Hayashida K., Iwasawa K., Kii T., Kunieda H., 
                 Makino F., Matsuoka M., 1995, Nature, 375, 695
\re Turner T.J., Pounds K.A.\ 1989, MNRAS  240, 833
\re Turner T.J. George I.M., Nandra K., Mushotzky R.F.\ 1997a, ApJS, 113, 23
\re Turner T.J. George I.M., Nandra K., Mushotzky R.F.\ 1997b, ApJ, 488, 164
\re Ueno S.\ 1997, Ph.D. Thesis, Kyoto University
\re Vignali C., Comastri A., Stirpe G.M., Cappi M., Palumbo G.G.C.,
                 Matsuoka M., Malaguti G., Bassani L.\ 1998, A\&A, in press
\re Vila-Costas M.B. \& Edmunds M.G.\ 1992, MNRAS  259, 121
\re Ward M.J., Wilson A.S., Penston M.V., Elvis M., Maccacaro T.,
                 Tritton K.P.\ 1978, ApJ  223, 788
\re Ward M., Penston M.V., Blades, J.C., and Turtle, A.J., 1980,
                   MNRAS, 193, 563
\re Warwick R.S., Pounds K.A.,Turner T.J.\ 1988, MNRAS  231, 1145 
\re Warwick R.S., Sembay S., Yaqoob T., Makishima K., Ohashi T., 
                 Tashiro M., Kohmura Y.\ 1993, MNRAS, 265, 412
\re Weaver K.A., Yaqoob T., Holt S., Mushotzky R.F., Matsuoka M., 
                 Yamauchi M.\ 1994, ApJ  436, L27
\re Weaver K.A., Yaqoob T., Mushotzky R.F., Nousek J., Hayashi I.,
                 Koyama K.\ 1997, ApJ  474, 675

\label{last}

\newpage
\markboth{}{}
\normalsize

\end{document}